\documentclass[12pt,preprint]{aastex}

\usepackage{emulateapj5}
\usepackage{aastexug}
\usepackage{amsmath}
\usepackage{graphicx}
\usepackage{subfigure}
\usepackage{apjfonts}

 \font\sevenrm=cmr7 scaled 1000

\def\gsim{\;\lower4pt\hbox{${\buildrel\displaystyle >\over\sim}$}\;}
\def\lsim{\;\lower4pt\hbox{${\buildrel\displaystyle <\over\sim}$}\;}
\def\grls{\;\lower4pt\hbox{${\buildrel\displaystyle >\over <}$}\;}

\begin{document}
%\title{Self-similar shocks in young stellar objects,
%H {\eightrm II} regions and Planetary Nebulae}
\title{Shocked similarity collapses and flows in star formation processes }

\shorttitle{Shocked Similarity Flows}

\shortauthors{Shen \& Lou}

\author{Yue Shen$^{1}$ and Yu-Qing Lou$^{1,2,3}$}

\affil{$^1$Physics Department and the Tsinghua Center for
Astrophysics (THCA), Tsinghua University, Beijing 100084, China;\\
$^2$Department of Astronomy and Astrophysics, The University
of Chicago, 5640 South Ellis Avenue, Chicago, IL 60637 USA;\\
$^3$National Astronomical Observatories, Chinese Academy of
Sciences, A20, Datun Road, Beijing 100012, China.}

%October 30, 2003 (Thursday) THCA
%November 18, 2003 (Tuesday) THCA
%November 23, 2003 (Sunday) Beijing home
%April 11, 2004 (Sunday) Beijing home
%April 12, 2004 (Monday) Beijing THCA
%April 13, 2004 (Tuesday) Beijing THCA
%April 14, 2004 (Wednesday) Beijing THCA
%April 15, 2004 (Thursday) Beijing THCA
%April 16, 2004 (Thursday) Beijing home; Mom's 75 Birthday.
%April 18, 2004 (Saturday) Beijing home; TziFan couple visit.
%April 19, 2004 (Monday) Beijing THCA.
%May   24, 2004 (Monday) Beijing THCA.
%July   5, 2004 (Monday) Beijing THCA.

\begin{abstract}
We propose self-similar shocked flow models for certain dynamical
evolution phases of young stellar objects (YSOs), `champagne
flows' of H {\!\sevenrm II} regions surrounding OB stars and
shaping processes of planetary nebulae (PNe). We analyze an
isothermal fluid of spherical symmetry and construct families
of similarity shocked flow solutions featured by: 1. either a
core expansion with a finite central density or a core accretion
at constant rate with a density scaling $\propto r^{-3/2}$;
%(a free-fall state);
2. a shock moving outward at a constant speed;
3. a preshock gas approaching a constant speed at large $r$
%(a wind or a breeze or an accretion flow)
with a density scaling $\propto r^{-2}$. In addition to
testing numerical codes, our models can accommodate diverse
shocked flows with or without a core collapse or outflow
and an envelope expansion or contraction.
%By adding physical aspects of asymmetry and/or magnetic
%field, our models can be further adapted for more realistic
%morphologies such as bipolar outflows in YSOs or PNe.
As an application, we introduce our model analysis
to observations of Bok globule B335.
\end{abstract}


\section{INTRODUCTION}

Far away from initial and boundary conditions, dynamical evolution
of a fluid may lead to self-similar phases with variable
profiles shape-invariant and magnitudes properly scaled (Sedov 1959).
%; Landau \& Lifshitz 1959).
%Similarity methods, which transform partial differential equations
%(PDEs) to ordinary differential equations (ODEs), greatly simplify
%nonlinear problems.
For star formation, self-similar solutions were
found to describe collapses of isothermal gas clouds (Larson
1969; Penston 1969; Shu 1977, hereafter S77). From an initial
time $t=0$ when the core becomes singular to the final stage
$t\rightarrow\infty$, Shu (1977) derived the
%so-called
`expansion-wave collapse solution' (EWCS) that has a free-fall
core of density $\rho\propto r^{-3/2}$, a radial infall speed
$u\propto r^{-1/2}$, a constant mass accretion rate at
$r\rightarrow 0$ and a static envelope of a singular isothermal
sphere (SIS) with the stagnation point moving outward at the sound
speed $a$.
%to enclose more infalling materials.
This EWCS scenario of `inside-out' collapses for star formation
has been advocated by Shu et al. (1987) and compared with
observations (e.g., Zhou et al. 1993, hereafter Z93; Choi et al.
1995; Saito et al. 1999, hereafter Sa99; Harvey et al. 2001).

Lou \& Shen (2004, hereafter LS04) re-examined this problem
and derived new solutions in the `semi-complete space'
($0<t<+\infty$), in contrast to the `complete space'
($-\infty<t<+\infty$) of Hunter (1977) and Whitworth \& Summers
(1985, hereafter WS). Gas flows at large $r$ together with the
two distinct asymptotic behaviors near the origin give rise to
infinitely many similarity solutions. Those solutions of
`envelope expansion with core collapse' (EECC) are especially
interesting.

Besides these shock-free solutions, Tsai \& Hsu (1995, hereafter
TH95) constructed a shocked self-similar collapse solution for a
situation where a central energy release initiates an outgoing
shock during a protostellar collapse of a low-mass star. Such a
shocked solution, matched with a static SIS envelope, has a
free-fall core yet with a lower mass accretion rate compared with
the EWCS. TH95 also constructed a shocked expansion solution of
a finite core density matched with a static SIS envelope, which
was recently generalized and subsumed into a family of shocked
`champagne flow' solutions by Shu et al. (2002, hereafter S02) for
expansions of H {\!\sevenrm II} regions around OB stars (Str\"omgren
1939; Tenorio-Tagle 1979, 1982; Newman \& Axford 1968; Mathews \&
O'Dell 1969; Franco, Tenorio-Tagle \& Bodenheimer 1990, hereafter FTB).

The main thrust of this Letter is to show that in the solution
framework of LS04, a variety of shocked similarity flow solutions
can be constructed and applied to astrophysical systems.
%Based on these solutions, we discuss astrophysical
%applications to YSOs, H {\!\sevenrm II} regions and PNe.

\section{SHOCKED SIMILARITY FLOW MODELS}

To study basic properties of shocked similarity flows
of spherical symmetry, we assume isothermality with a
constant sound speed $a$.
%Isothermality is well satisfied in most star-forming and H
%{\!\sevenrm II} regions but may be violated in the PNe regime.
%Furthermore we assume spherical symmetry.
The standard self-similar nonlinear ordinary
differential equations (ODEs) (S77;
%{\it Hunter 1977; WS;}
TH95; S02; LS04) are:
\begin{equation}
[(x-v)^2-1]\frac{dv}{dx}
=\bigg[\alpha(x-v)-\frac{2}{x}\bigg](x-v)\ ,
\end{equation}
\begin{equation}
[(x-v)^2-1]\frac{1}{\alpha}\frac{d\alpha}{dx}
=\bigg[\alpha-\frac{2}{x}(x-v)\bigg](x-v)\ ,
\end{equation}
\begin{equation}
m=x^2\alpha(x-v)\ ,
\end{equation}
where the independent similarity variable $x\equiv r/(at)$
and $\alpha(x)$, $v(x)$, $m(x)$ are the reduced density,
radial speed and enclosed mass, respectively. The physical
density $\rho$, radial speed $u$ and enclosed mass $M$ are
obtained by the similarity transformation
\begin{equation}
\begin{split}
\rho(r,t)=\frac{\alpha(x)}{4\pi Gt^2}\ ,\ \ \ \
u(r,t)=av(x)\ ,\ \ \ \
M(r,t)=\frac{a^3t}{G}m(x)\ .
\end{split}
\end{equation}
The analytical asymptotic solutions are (i) for $x\rightarrow +\infty$,
\begin{eqnarray}
v\rightarrow V\ , \qquad \alpha\rightarrow A/{x^2}\ ,\
\qquad m\rightarrow Ax\ ,
\end{eqnarray}
where $V$ and $A$ are two parameters;
and (ii) for $x\rightarrow 0$, either
\begin{equation}
v\rightarrow-(2m_0/x)^{1/2},\ \ \
\alpha\rightarrow[m_0/(2x^3)]^{1/2},
\ \ \  m\rightarrow m_0\ ,
\end{equation}
or
\begin{equation}
v\rightarrow 2x/3 \ , \qquad \ \alpha\rightarrow B\ ,
\qquad \  m\rightarrow Bx^3/3\ ,
\end{equation}
where $m_0$ and $B$ are two constant parameters (e.g., LS04).

There exist two exact solutions, viz., the SIS solution (Bonner
1956; Chandrasekhar 1957) and the homogeneous `Hubble flow'
solution (WS; S02). ODEs $(1)-(3)$ can be solved numerically
to connect proper asymptotic solutions $(5)-(7)$.

The isothermal shock conditions
\footnote{The two jump conditions in the shock
framework are the mass and momentum conservations
$\rho_d(u_d-u_s)=\rho_u(u_u-u_s)$ and
$a_d^2\rho_d+\rho_du_d(u_d-u_s)=a_u^2\rho_u+\rho_uu_u(u_u-u_s)$,
where $u_s$ is the shock speed.}
are the mass conservation
$$\alpha_d(v_d-x_{sd})a_d=\alpha_u(v_u-x_{su})a_u$$
and the momentum conservation
$$\alpha_d[1+v_d(v_d-x_{sd})]a_d^2=\alpha_u[1+v_u(v_u-x_{su})]a_u^2$$
with the energy conservation involving radiative losses (e.g., Courant
\& Friedrichs 1976; Spitzer 1978), where subscript $d$ ($u$) denotes
the downstream (upstream) of a shock, $a_dx_{sd}=a_ux_{su}$ is the
shock speed, $a_d$ and $a_u$ are the downstream and upstream isothermal
sound speeds, and $x_{d}=r/(a_dt)$ and $x_{u}=r/(a_ut)$ are the downstream
and upstream independent similarity variables. For a sound speed ratio
$\tau\equiv a_d/a_u$ with $\tau x_{sd}=x_{su}$, we have
\begin{eqnarray}
v_d-x_{sd}-\tau(v_u-x_{su})=(\tau v_d-v_u)(v_u-x_{su})(v_d-x_{sd})\ ,
\nonumber
\end{eqnarray}
\begin{equation}
\qquad\qquad\alpha_d/\alpha_u=(v_u-x_{su})/[\tau (v_d-x_{sd})]\ .
\qquad\qquad\qquad\
\end{equation}
The special isothermal shock conditions of $\tau=1$ with
$x_{sd}\equiv x_{su}=x_s$ (TH95; S02) are adopted in our
numerical examples.
%with the same upstream and downstream sound speeds.
%We follow this idea and set $x_{sd}\equiv x_{su}=x_s$.

%Note there are two possible asymptotic behaviors when
%$x\rightarrow 0$. For asymptotic behaviors (6),
We take the so-called type 2 solutions (Hunter 1977, 1986; LS04)
that cross the sonic critical line $x-v=1$ once at $x_{*}<1$ and
approach asymptotic diverging behavior (6) for $x\rightarrow 0$
as downstream (post-shock) flows. By integrating a type 2
solution toward the sonic critical line, imposing shock condition
(8) at each integration step and continuing further toward
$x\rightarrow +\infty$ (upstream), we construct a family
of shocked similarity flows displayed in Fig. 1 with $\tau=1$.
%upstream solutions corresponding
%to the initial downstream solution.
The resulting upstream solutions approach
asymptotic behavior (5) as $x\rightarrow +\infty$.
%Several this kind of shocked
%similarity solutions are shown in Fig. 1.
Such a solution is uniquely determined by two parameters: the
central mass accretion rate $m_0$ (or equivalently, the crossing
point $x_{*}<1$) and the shock location $x_s$. Simultaneously
determined are the velocity and density parameters $V$ and $A$ as
$x\rightarrow +\infty$ in (5). For each such solution (denoted as
Class I), a core collapses with a constant mass accretion rate
$m_{0}a^3/G$ and a shock travels at speed $ax_s$ outward to match
with either an expanding or a contracting envelope; at very large
$r$, the flow approaches a constant speed as either wind
(including breeze and SIS) or inflow -- all with density
scalings of $\propto r^{-2}$ (LS04).

For downstream solutions with asymptotic behavior (7), we can
construct Class II solutions of shocked flows in parallel with
the Class I solutions. Such a solution is uniquely determined by
the finite core parameter $B$ and the shock location $x_s$. For
each Class II solutions, a core expands
%(since $v\rightarrow 2x/3$ for $x\rightarrow 0$)
and a shock travels outward at a speed $ax_s$ to match with
envelope expansion or contraction of a constant speed $V\neq 0$ or
a breeze $V=0$ (S02) as $x\rightarrow +\infty$ shown by heavy
solid curves in Fig. 1.

%%\clearpage

\begin{center}
\footnotesize
\begin{tabular}{ccccc}\hline
%\hline
 Type   &  Description             & Shock location $x_s$  &  $V$  & $A$  \\ \hline
Class I &  $m_0=0.406$, $x_*=0.23$ & 0.43                  &-2.495 & 0.250\\
        &                          & 1.03                  &-1.126 & 0.847\\
        &                          & 1.37                  & 0.0   & 1.865\\
        &                          & 1.43                  & 0.235 & 2.158\\
Class II&  $B=0.1$                 & 1.30                  &-2.192 & 0.0263\\
        &                          & 2.00                  &-0.711 & 0.148\\
        &                          & 2.80                  &0.637  & 0.640\\
\hline
\end{tabular}
%%\tablecaption{Parameters for example shocked similarity
%%solutions.}
\end{center}

\begin{deluxetable}{ccccc}
\tablewidth{0pt} \tabletypesize{\small} \tablecaption{Parameters
for example shocked similarity solutions.} \tablecolumns{5}
\tablehead{Type  & Description & Shock location $x_s$ & $V$ & $A$
} \startdata
Class I &  $m_0=0.406$, $x_*=0.23$ & 0.43                  &-2.495 & 0.250\\
        &                          & 1.03                  &-1.126 & 0.847\\
        &                          & 1.37                  & 0.0   & 1.865\\
        &                          & 1.43                  & 0.235 & 2.158\\
Class II&  $B=0.1$                 & 1.30                  &-2.192 & 0.0263\\
        &                          & 2.00                  &-0.711 & 0.148\\
        &                          & 2.80                  &0.637  & 0.640\\
\enddata
\end{deluxetable}

%%\clearpage

A few important features are noted here. 1. Shocked similarity
solutions of Class I qualitatively share properties of the EWCS
(inside-out collapse) at small $r$ with density $\rho\propto
r^{-3/2}$, infall speed $u\propto r^{-1/2}$ and constant core mass
accretion rate $m_0$. 2. The main distinction is that the upstream
solutions for both classes can be an inflow or a wind (LS04) or a
breeze (S02) or a SIS\footnote{There are infinitely many discrete
downstream solutions with extremely large $B$ or small $m_0$ and
oscillating $v(x)$ around zero to drive shocks into a SIS envelope.}
(TH95). 3. For very large $B$ or small $m_0$ (or small $x_*$), the
downstream solution $v(x)$ can oscillate (Hunter 1977 for the
former and LS04 for the latter). The core mass accretion rate
can be extremely low for the latter.
%3. It is possible that the upstream solution is the static
%SIS as adopted by TH95 or approaches a breeze solution with
%$V=0$ for $x\rightarrow +\infty$ as in S02.

%%\clearpage

\vglue 0.2cm \figurenum{1} \centerline{
\includegraphics[scale=0.5]{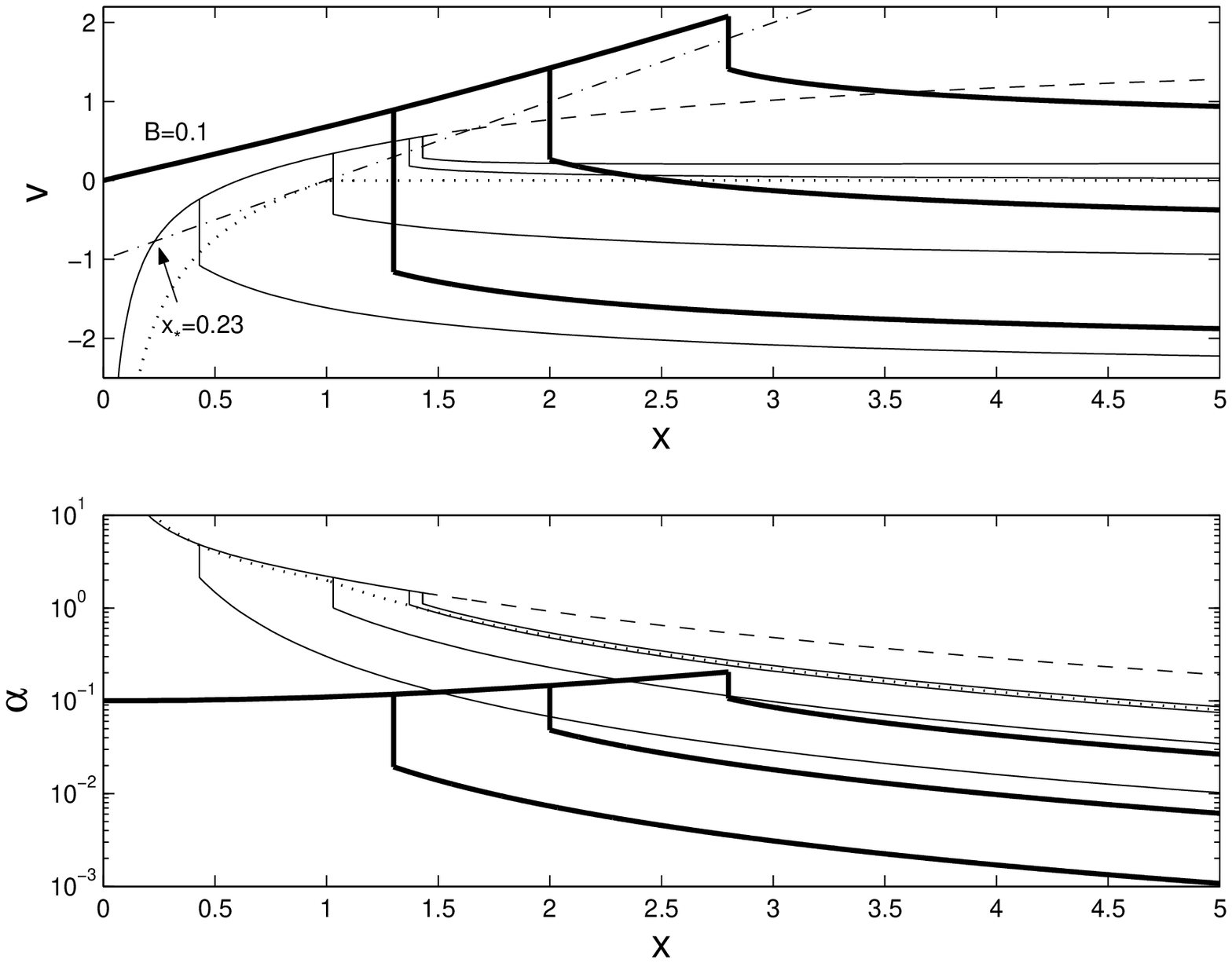}}
\figcaption{Examples of Classes I and II (light and heavy solid
curves) shocked flow solutions. For Class I solutions, we choose
the downstream solution to cross the sonic critical line
$x-v=1$ [the dash-dotted line for $v(x)$] at $x_{*}=0.23$ with
%a core mass accretion rate
$m_0=0.406$. An infinite number of solutions can be constructed
for various shock locations $x_s$ to match asymptotic behavior (5)
as $x\rightarrow +\infty$ with various $V$ and $A$ (either a wind
or a breeze or a contraction). This special
downstream solution, referred to as the CP1 EECC solution in LS04,
can analytically cross the sonic critical line again at
$x_*(2)=1.65$ (dashed line). For Class II solutions in parallel,
we choose the downstream solution of $B=0.1$ and various $x_s$ to
match (5) with various $V$ and $A$.
%(either a wind, or a breeze, or a SIS, or a contraction).
%It also fans out infinitely many downstreams at various shock location
%$x_s$ and approach asymptotic behavior (5) when $x\rightarrow +\infty$.
Parameters of these examples are summarized in Table 1.
The dotted lines stand for the EWCS.} \vglue 1cm

%%\clearpage

%\begin{center}
%\footnotesize
%\begin{tabular}{ccccc}\hline
%\hline
% Type   &  Description             & Shock location $x_s$  &  $V$  & $A$  \\ \hline
%Class I &  $m_0=0.406$, $x_*=0.23$ & 0.43                  &-2.495 & 0.250\\
%        &                          & 1.03                  &-1.126 & 0.847\\
%        &                          & 1.43                  & 0.235 & 2.158\\
%Class II&  $B=0.1$                 & 1.30                  &-2.192 & 0.0263\\
%        &                          & 2.00                  &-0.711 & 0.148\\
%        &                          & 2.80                  &0.637  & 0.640\\
%\hline
%\end{tabular}
%\end{center}

\section{STAR FORMATION PROCESSES}

\subsection{Protostellar Collapses and Outflows (Class I solutions)}

Regarding the EWCS scenario for star formation (Shu et al.
1987), TH95 argued that a central energy output during a
protostellar collapse can initiate a shock into a SIS
envelope with a lower core mass accretion rate.
%S02 enriched this scenario by allowing a shock into various
%envelope breezes [$v(x)\rightarrow 0$ as $x\rightarrow\infty$].
Based on solutions of LS04, we envision a much more general
scenario for a central inside-out collapse with an adjustable
central mass accretion rate and with an outgoing shock into
various possible envelope flows such as winds, breezes, inflows
or SIS at large $r$.
%While TH95 limited their consideration to a SIS envelope, and
%hence missed lots of other shocked solutions which may possess
%various core mass accretion rate as well as non-vanishing fluid
%velocity at large radii.
The core mass accretion rate is $\dot{M}(0,t)=m_0a^3/G$.
%with $m_0$ as an accretion parameter.
While the EWCS has $m_0=0.975$ and the example of TH95 has
$m_0=0.105$, our Class I solutions for shocked flows give
rise to a continuous range of $0<m_0<0.975$.
%Suppose the shock location is $x_s$, then the shock front travels
%outward at constant speed $x_sa$. With specified $m_0$ and $x_s$,
%the constant velocity and $r^{-2}$ density profile at large radii
%are also determined. Such
These solutions can describe certain similarity evolution phases
during protostellar collapses and outflows. With these physical
concepts, we now turn to the well-observed protostellar cloud
B335 that possesses a collapsing core.

Besides the observed bipolar CO outflows on large scales, a SIS
environment is presumed to surround the central collapse region
and has been modeled using the EWCS (S77; Z93; Sa99). From a
temperature profile estimated from dust emissions (Zhou et al.
1990), B335 globule was approximated as an isothermal cloud with
an effective sound speed $a_{\hbox{eff}}\sim 0.23$ km s$^{-1}$. In
essence, a gross spherical symmetry is assumed for B335 cloud by
regarding the bipolar outflows as additional features produced by
a circumstellar disk within $\lesssim100$ AU (Harvey et al. 2003).
In the EWCS modeling (Z93), a core collapses as a free fall while
an envelope remains static with an estimated infall radius
$r_{\hbox{inf}}\sim 0.03$ pc.
%{\bf A density `kink' from
%$\rho\propto r^{-2}$ to $\rho\propto r^{-3/2}$ might have been
%seen around $r_{\hbox{inf}}$ as predicted by the EWCS model.}

Given some successes of the EWCS model in interpreting the density
and velocity profiles (via simulations of spectral lines; Z93;
Choi et al. 1995; Velusamy et al. 1995), departures from the EWCS
model exist. In Sa99, the estimated infall speed at $r\sim 2200$
AU is $\sim 0.14$ km s$^{-1}$ for $r_{\hbox{inf}}\sim 7200$ AU,
smaller than the EWCS prediction of $\sim 0.34$ km s$^{-1}$ by a
factor of $\sim 2.4$, and the estimated mass accretion rate is
$1.2-2.4\times10^{-6}M_\odot$ yr$^{-1}$, corresponding to $m_0\sim
0.418-0.836$ that tends to be less than the EWCS prediction
$m_0=0.975$ (Z93). Meanwhile, the column density of the outer
envelope at $r=10,000$ AU is $N\sim 6.3\times10^{21}$cm$^{-2}$,
slightly larger than the EWCS prediction $N\sim 5.4\times10^{21}$
cm$^{-2}$ (Sa99). This last difference is aggravated by the
near-IR extinction study of Harvey et al. (2001) that gives an
envelope density $3-5$ times higher than EWCS prediction while
the density $\rho$ scales as $r^{-2}$. HCN spectral analysis of
B335 seems to indicate a central core collapse with an envelope
expansion distinctly different from bipolar CO outflows.
%(A. Lapinov 2004, private communications).

The major advantage of our shocked similarity flow model is
the capability of accommodating observations of B335 globule
%(though with uncertainties)
while making testable predictions for envelope expansion in
contrast to a SIS of the EWCS. As an example, we introduce the
unshocked CP1 EECC solution (dashed curves in Fig. 1) that
analytically passes the sonic critical line twice (LS04). While
the inner core freely falls with a $\rho\propto r^{-3/2}$, the
envelope expands with a $\rho\propto r^{-2}$. In contrast to the
EWCS, the transition across $r_{\hbox{inf}}$ is smooth without a
`kink'. Still taking $x=1$ for $r_{\hbox{inf}}=7200$ AU (or an age
of the similarity process is $\sim 1.5\times10^5$ yr in the EWCS
model, Z93), the infall speed at $r\sim 2200$ AU is $\sim 0.12$ km
s$^{-1}$ and the mass accretion rate given by CP1 EECC solution is
$m_0=0.406$ less than a half of 0.975 in the EWCS.
%{\it These are in much better agreement with
%observational inferences of Saito et al. (1999)}.
At large $r$, the density approaches $\rho\sim
Aa_{\hbox{eff}}^2/(4\pi G r^2)$ with $A=5.158$ (Table 3 of LS04).
That is, our CP1 EECC model gives an envelope density $\sim 2.5$
times higher than that of a SIS in the EWCS (i.e. $A=2$). This
tends to agree with the inference of $3-5$ times by Harvey et al.
(2001). One distinctive consequence of our isothermal model is
that the envelope, instead of a SIS, expands with a
speed\footnote{In reality, a polytropic model should be more
pertinent as discussed in LS04. For a polytropic gas, the
asymptotic wind speed should tend to vanish at large $r$.} of
$\sim 0.25$ km s$^{-1}$ at $r=25,000$ AU. This is qualitatively
consistent with the preliminary HCN spectral analysis (A. Lapinov
2004, private communications). By inserting a shock at a proper
location, it has a range to adjust the data fit for density and
speed profiles and there is a preliminary evidence of shock in
B335 system\footnote{Practically,
%as also pointed out by the referee,
it might be far easier to detect the presence of large-scale
shocks than to estimate the velocity and density profiles as
shocks will leave non-kinematic signatures amenable to
observations, i.e., through emission diagnostics from highly
compressed post-shock gas or chemical or radiative cooling
processes.} (e.g., Nisini et al. 1999).
%Solutions other than CP1 may fit the data even better with
%pertinent density and velocity distributions for both inner
%core and outer envelope, and shocks may explain some
%observed specific line emissions (Nisini et al. 1999).
Sharing the qualitative features of the EWCS in the central
region, our Class I solutions (including CP1 EECC solution)
of shocked similarity flows are versatile to model other
protostellar cloud or Bok globule systems.

%Another aspect in this field is the commonly observed outflows
%in younger YSOs, which are often bipolar or collimated jets
%(Bachiller 1996; Feigelson \& Montmerle 1999). Hydrodynamical or
%MHD-disk models are built to explain such phenomena (Raga \&
%Cant\'{o} 1989; K\"{o}nigl 1989; Frank \& Mellema 1996). Now
%within the hydrodynamical regime, we believe our generalized
%self-similar shock model is adaptable because it contains both
%the collapse part which continuously feeds the protostar and the
%expansion part which powers the outflows as well as shocks. When
%the expanding part suffers the outer aspherical density profile
%(i.e., a dense toroid due to the overall rotation of the cloud),
%bipolar outflows are created (in some sense like the GISW model
%in PNe shaping, e.g. Kwok et al. 1978; Balick \& Frank 2002 and
%reference therein), although it may not be applicable in the B335
%system since the bipolar outflow is believed to caused by  a well
%collimated driving wind from the circumstellar disk (Chandler \&
%Sargent 1993; Harvey et al. 2003).

%%\clearpage

\figurenum{2} \centerline{
\includegraphics[scale=0.5]{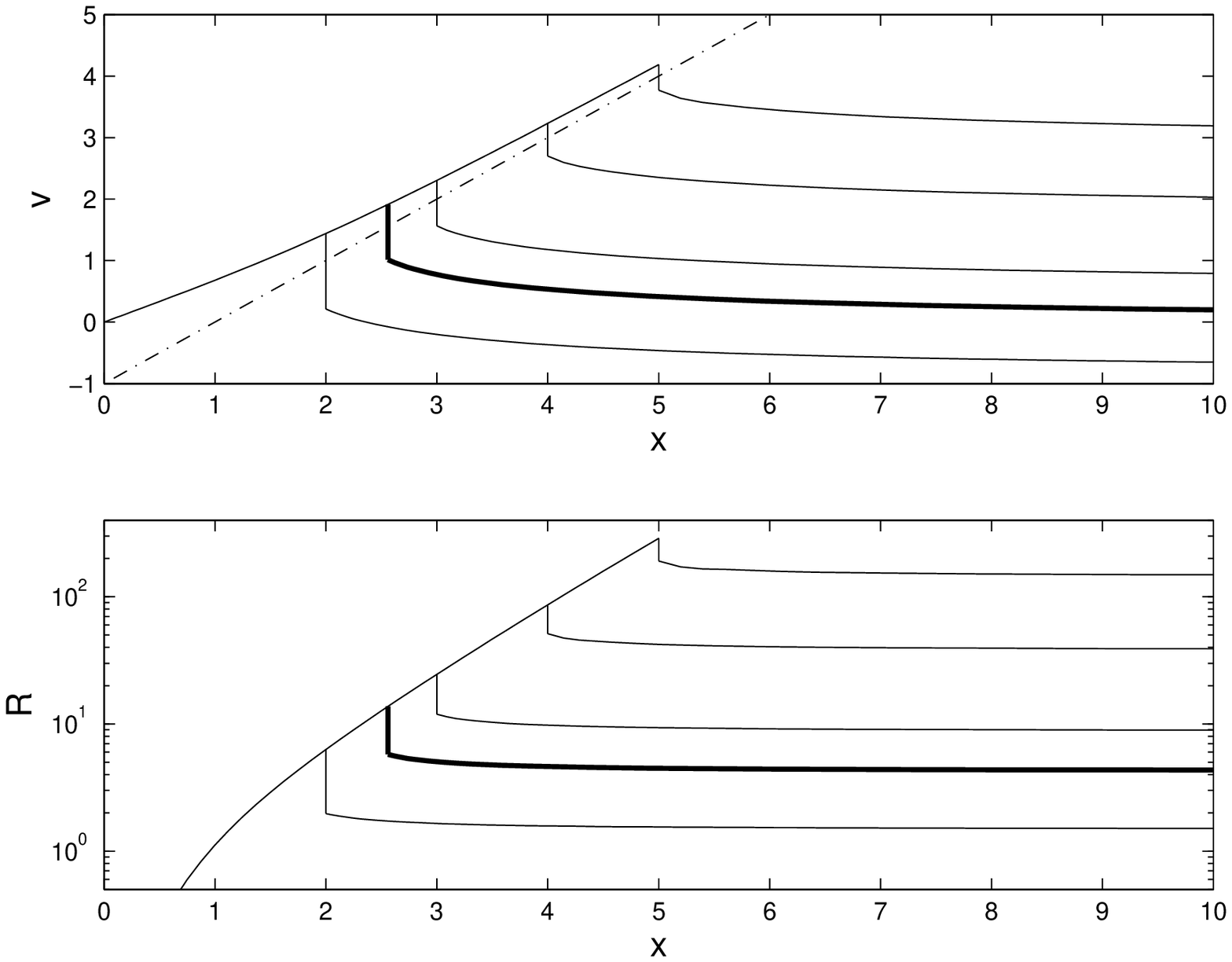}}
\figcaption{The family of Class II solutions with an invariant
downstream (S02) as $B\rightarrow 0$ ($B=10^{-6}$). Here, $\alpha(x)$
is rescaled as $R(x)\equiv x^2\alpha(x)/B$. The heavy solid line
stands for the `champagne breeze' of S02. In addition to breezes and
contractions at large $r$, many shocked similarity flows approach
constant wind speeds as $x\rightarrow\infty$ that can be either
supersonic or subsonic.} \vglue 1cm

%%\clearpage

\subsection{`Champagne Flows' in H {\!\sevenrm II}
Regions (Class II solutions)}

As massive OB stars form in molecular clouds, intense ultraviolet
radiations ionize surrounding H {\!\sevenrm I} clouds and carve out
H {\!\sevenrm II} regions (Str\"omgren 1939). As an ionization front
passes by, a pressure gradient develops between H {\!\sevenrm II}
and H {\!\sevenrm I} regions to drive flows. Shocks can naturally
occur in such an impulsive expansion phase. An expansion of a
H {\!\sevenrm II} region embedded in a uniform ambient medium has
been studied (e.g., Newman \& Axford 1968; Mathews \& O'Dell 1969).
Other possible environs may involve power-law cloud density profiles
of $\propto r^{-n}$ with $1\lsim n\lsim 3$ and a mean of $n\cong 2$
(e.g. Arquilla \& Goldsmith 1985). As ionization or shock fronts
encounter a strong negative
density gradient, the ensuing supersonic expansion of the ionized
gas is referred to as the `champagne phase'\footnote{An ionization
front may run into edges of a molecular cloud and burst into a
low-density intercloud medium to drive `champagne flows'
(Tenorio-Tagle 1979).} (FTB). When a fast moving ionization front
cannot be confined under certain conditions (FTB) and ionizes the
entire cloud in a relatively short time, the inevitable expansion
of H {\!\sevenrm II} regions might evolve into a self-similar
phase (S02).

In the scenario above, a cloud is presumed static initially and
the H {\!\sevenrm II} region forms instantaneously at $t=0$ when
the stellar nuclear burning turns on. Our more general scenario
would allow a molecular cloud with asymptotic flows (either
winds and breezes or accretions and contractions) at large $r$.
%Processes prior to stellar nuclear ignition
%allow various conceivable possibilities.
To model expansions of H {\!\sevenrm II} regions as the ionization
front travels, we focus on various possible large-scale systematic
flows at large $r$ and put aside messy central star-forming
processes on much smaller scales. Within the ionized cloud sphere,
asymptotic winds and accretions at large $r$ can be connected to
inner champagne flows via transonic shocks. As transients peter out,
an expanding H {\!\sevenrm II} region may become self-similar with
an outgoing shock to match with an asymptotic wind, breeze or accretion.

%%\clearpage
%
%\figurenum{2} \centerline{
%\includegraphics[scale=0.5]{f2.eps}}
%\figcaption{The family of Class II solutions with an invariant
%downstream (S02) as $B\rightarrow 0$ ($B=10^{-6}$). Here, $\alpha(x)$
%is rescaled as $R(x)\equiv x^2\alpha(x)/B$. The heavy solid line
%stands for the `champagne breeze' of S02. In addition to breezes and
%contractions at large $r$, many shocked similarity flows approach
%constant wind speeds as $x\rightarrow\infty$ that can be either
%supersonic or subsonic.} \vglue 1cm
%
%%\clearpage

The shocked breeze solutions has been used to model
similarity expansions of H {\!\sevenrm II} regions (S02).
%{\bf In the regime of Class II solutions,}
For each `champagne breeze' solution of S02, we can use the
same downstream to construct Class II solutions with shocked
upstream flows which approach breezes, winds and accretions
at large $r$. As $B\rightarrow 0$, the downstream solution
becomes invariant (S02) and the resulting family of upstream
solutions is displayed in Fig. 2.

Observations do suggest the presence of systematic flows in H
{\!\sevenrm II} regions and models involving stellar winds were
proposed (e.g. Dyson 1977; Hippelein \& M\"unch 1981; Comer\'on
1997). Such outflows may be supersonic of a few $\hbox{km\
s}^{-1}$ (e.g., Rela\~{n}o et al. 2003).
%{\bf Meanwhile, shocks can be detected through intense
%emissions from highly compressed post-shock gas.}
In reference to our Class II models, shocked
similarity `champagne winds' in H {\!\sevenrm II} regions
distinctly differ from `champagne breezes' or SIS by having both
constant wind speeds at large $r$ and outgoing shocks at constant
speed. Compared with breezes, `champagne winds' would dynamically
impact the surrounding interstellar medium more efficiently.

\subsection{Implications for Planetary Nebulae (PNe) }

%Our isothermal model fails in some systems such as PNe (proto-PNe)
%and supernova remnants (SNRs). Nevertheless, it is easy to extend
%the isothermal model to the polytropic case (LS04), which seems to
%give qualitatively the similar results (Lou \& Gao?). Therefore we
%use it to grossly describe the PNe and SNRs systems.

In hydrodynamic models for shaping PNe, the interacting stellar
wind (ISW) scenario (Kwok et al. 1978) is often invoked. In a few
spherical cases, a much faster stellar wind from the core catches
up with a slower dense wind -- the remnant of the AGB phase, and
form shocks. Our shocked similarity flow solutions of Class I
clearly show the physical feasibility that the shaping of a PNe
involving ISW can be concurrently accompanied by a central
accretion or infall towards a proto white dwarf in a similarity
manner. In order to model bipolar PNe other than round or mildly
elliptical morphologies, aspherical aspects are included in the
so-called generalized ISW (GISW) model (Balick \& Frank 2002).
%{\bf In the framework of GISW, Dwarkadas et al. (1996) found
%similarity solutions for the evolution of shocked shells.}
By flow
instabilities, a shocked envelope can expand to create various
nebula morphologies while the central core continuously accretes
materials to form a white dwarf. As a proto white dwarf
continuously accretes materials to approach the Chandrasekhar
limit of $\sim 1.39M_{\odot}$, there might be even a
possibility of igniting a type Ia supernova explosion.
%Similar processes may occur in SNRs where the ejecta falls
%back and forms fallback disks around the central neutral star.

%\section{summary and discussions}
%We have modelled self-similar processes where large-scale flows
%involving core collapse, envelope expansions or contractions, as
%well as shocks can well coincide. The isothermal spherical shock
%model has been applied in various circumstances. The plenty and
%simplicity of our solutions provide useful estimations for testing
%numerical codes and offer theoretical models for fitting with
%observations in the dynamics and kinematics of YSOs, H {\!\sevenrm
%II} regions, PNe or SNRs using proper fluid description.

\section*{Acknowledgments}
This research has been supported in part by the ASCI Center
for Astrophysical Thermonuclear Flashes at the U. of Chicago
under DoE contract B341495, by the Special Funds for
%Major State Basic Science Research Projects
MSBSRP of China, by the THCA, by the
Collaborative Research Fund from the
%National Natural Science Foundation
NSF of China (NSFC) for Young Outstanding Overseas
Chinese Scholars (NSFC 10028306) at the NAOC,
%Chinese Academy of Sciences
CAS, by NSFC grant 10373009 at the Tsinghua U., and by the
Yangtze Endowment from the MoE at the Tsinghua U.. Affiliated
institutions of Y.Q.L. share this contribution.

\end{document}